\documentclass[aps, prl, reprint, superscriptaddress]{revtex4-1}

\usepackage{graphicx}
\usepackage{dcolumn}
\usepackage{amsmath}
\usepackage{amssymb}
\usepackage{color}
\usepackage{bm}
\usepackage{braket}
\usepackage{txfonts}
\usepackage[colorlinks=true, letterpaper=true, pdfstartview=FitV, linkcolor=blue, citecolor=blue, urlcolor=blue]{hyperref}

\newcommand{\RNum}[1]{\uppercase\expandafter{\romannumeral #1\relax}}
\newcommand{\N}{\mathcal{N}}
\newcommand{\U}{\mathcal{U}}
\newcommand{\Tr}{\mathrm{Tr}}

\newcommand{\RH}{\mathrm{H}}
\newcommand{\RL}{\mathrm{L}}
\newcommand{\RI}{\mathrm{I}}
\newcommand{\RT}{\mathrm{T}}

\begin{document}

\title{Digital Quantum Simulation of the Lindblad Master Equation and Its Nonlinear Extensions via Quantum Trajectory Averaging}

\author{Yu-Guo Liu}
\affiliation{Beijing National Laboratory for Condensed Matter Physics, Institute of Physics, Chinese Academy of Sciences, Beijing 100190, China}
\affiliation{Center on Frontiers of Computing Studies, School of Computer Science, Peking University, Beijing 100871, China}
\author{Heng Fan}
\affiliation{Beijing National Laboratory for Condensed Matter Physics, Institute of Physics, Chinese Academy of Sciences, Beijing 100190, China}
\affiliation{School of Physical Sciences, University of Chinese Academy of Sciences, Beijing 100049, China}
\affiliation{Beijing Academy of Quantum Information Sciences, Beijing 100193, China}
\affiliation{Hefei National Laboratory, Hefei 230088, China}

\author{Shu Chen}
\email{schen@iphy.ac.cn}
\affiliation{Beijing National Laboratory for Condensed Matter Physics, Institute of Physics, Chinese Academy of Sciences, Beijing 100190, China}
\affiliation{School of Physical Sciences, University of Chinese Academy of Sciences, Beijing 100049, China}

\date{\today}

\begin{abstract}
Since precisely controlling dissipation in realistic environments is challenging, digital simulation of the Lindblad master equation (LME) is of great significance for understanding nonequilibrium dynamics in open quantum systems. However, achieving long-time simulations for complex systems with multiple dissipation channels remains a major challenge, both theoretically and experimentally. Here, we propose a 1-dilation digital scheme for simulating the LME based on quantum trajectory averaging without postselection. By rigorously matching the stochasticity inherent in quantum trajectories with the probabilistic outcomes of quantum measurements, our method effectively translates the classically established quantum jump algorithm into executable quantum circuits. A key advantage of our method is that it overcomes the exponential suppression of success probability seen in some existing postselection-dependent schemes, especially for long-time evolution or systems with numerous jump operators. Moreover, the scheme can be extended to a 2-dilation framework for the nonlinear LME with postselection, bridging the full LME and non-Hermitian Hamiltonian dynamics. This extended scheme provides a digital approach for exploring the interplay between non-Hermitian Hamiltonians and dissipative terms within a monitored quantum dynamics framework.
\end{abstract}

\maketitle

%%%%%%%%%%%%%%%%%%%%%%%%%%%%%%%%%%%%%%%%%%%%%%%%%%%%%%%%%%%%%%%%%%%%%%%%%%%%%%%%%%%%%%%%%%%%%%%%%%%%%%%%%%%%%%%%%%%%%%%%
%%%%%%%%%%%%%%%%%%%%%%%%%%%%%%%%%%%%%%%%%%%%%%%%%%%%%%%%%%%%%%%%%%%%%%%%%%%%%%%%%%%%%%%%%%%%%%%%%%%%%%%%%%%%%%%%%%%%%%%%
\emph{ \color{blue} Introduction.---}
The Lindblad master equation (LME) plays a crucial role in the study of open quantum systems, providing a rigorous framework to describe the quantum system coupling with the environment under the Born-Markov approximation~\cite{Lindblad1976,Gardiner1985,Daniel2020}. In recent years, the demand for precise control over dissipative processes has driven extensive theoretical efforts to explore novel nonequilibrium dynamical phenomena within the Lindblad framework, such as Liouvillian flat bands~\cite{Hugo2023}, postselected skin effects~\cite{Hugo2025}, and localization in open quantum systems~\cite{Yusipov2017,Vakulchyk2018,YLiu2024}. These studies often require carefully engineered dissipation operators, which are difficult to implement on analog simulation platforms. In contrast, digital quantum platforms provide a standardized and promising approach to simulation.

The primary challenge of digitally simulating the LME lies in implementing its intrinsically non-unitary nature using unitary logic gates. This difficulty can be addressed through approaches such as dilation methods~\cite{RCleve2017,ZDing2024,ZHu2020,KMarsden2021,ZHu2022,ASchlimgen2021,ASchlimgen2022,NSuri2023,DMazziotti2022,EOh2024,SPeng2025,LHu2019,JHan2021,WJYu2025}, which effectively embed non-unitarity into enlarged unitary gates, and hybrid quantum-classical algorithms, including variational quantum simulation~\cite{XYuan2019,SEndo2020,Mahdian2020,JLuo2024} and the quantum imaginary time evolution method~\cite{MMotta2020,HKamakari2022}. However, for large systems, numerous dissipation sources, and long-time evolution, significant theoretical and experimental challenges remain. For instance, probabilistic simulation approaches relying on postselection, such as those based on Sz.-Nagy dilation (SND)~\cite{ZHu2020,KMarsden2021,ZHu2022}, the linear combination of unitaries (LCU) method~\cite{ASchlimgen2021,ASchlimgen2022,NSuri2023}, and singular-value decomposition (SVD)~\cite{DMazziotti2022,EOh2024}, face the challenge of a rapidly diminishing overall success probability as the number of time steps or dissipation sources increases. For methods that avoid postselection, such as the one proposed by Ding \textit{et al.}~\cite{ZDing2024} based on the Stinespring dilation of Kraus operators, the critical issue often lies in the frequent requirement for larger number of auxiliary qubits, which can potentially affect the scalability of the simulation, especially on near-term quantum hardware with limited qubit counts. Recently, Peng \textit{et al.}~\cite{SPeng2025} proposed a trajectory-inspired Lindbladian simulation that reduces the circuit complexity with respect to the number of jump operators. However, it still relies on oblivious amplitude amplification to overcome the postselection problem, which increases the circuit depth and the required control precision.

\begin{figure}[htbp]\centering
\includegraphics[width=8.5cm]{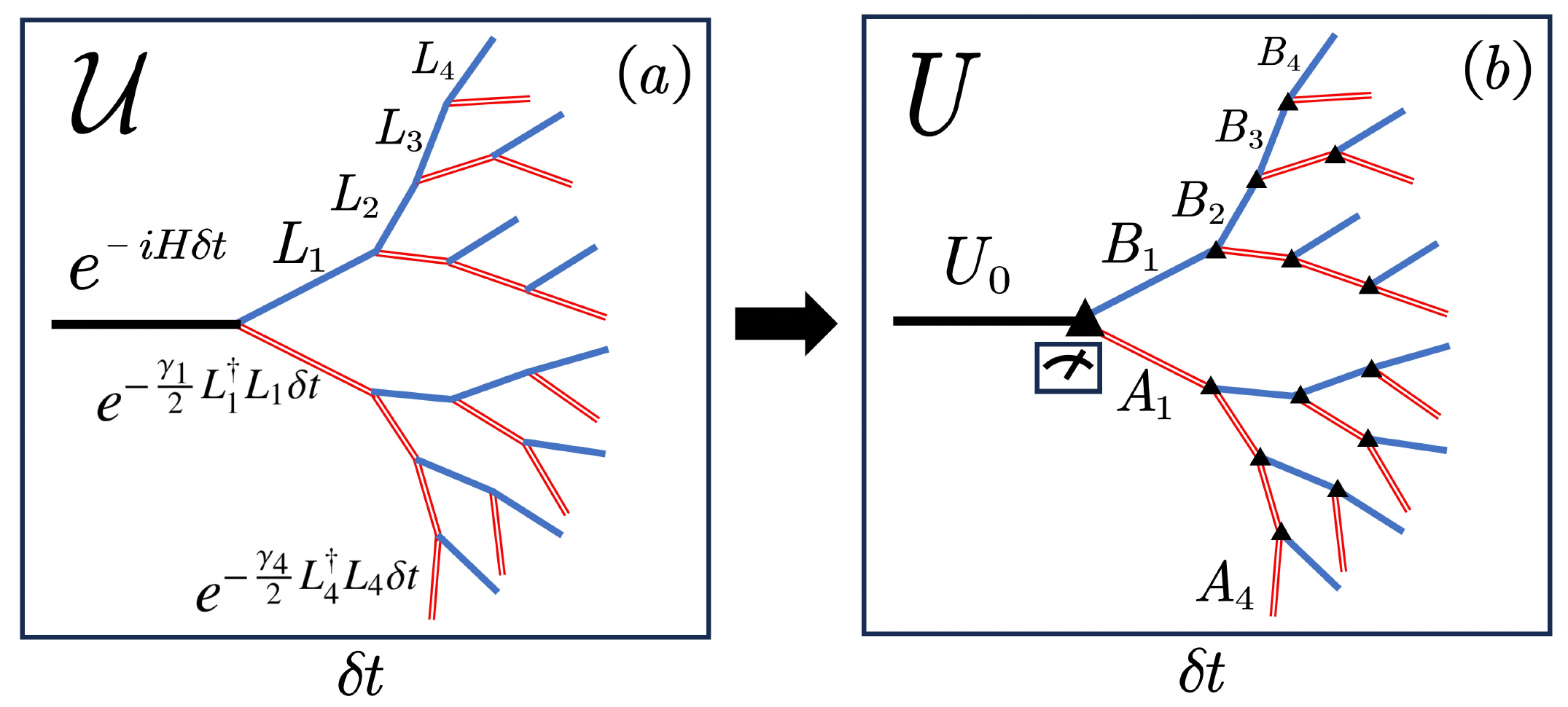}
\caption{ Realizing stochastic quantum trajectories via quantum circuit measurements. (a) In the classical algorithm, the system undergoes Hamiltonian evolution (black) followed by a stochastic choice between non-unitary evolution (red double lines) and quantum jumps $L_\mu$ (blue). The branching probabilities are determined by Eq.~(\ref{tadt}). (b) In the circuit implementation, the unitary $U_0$ corresponds to the Hamiltonian part, while the operators $A$ and $B$ (Eq.~(\ref{ABC})) represent the non-unitary and jump components. Ancilla measurements (black triangles) facilitate the stochasticity for LME, while NLME implementation further requires postselection on the measurement outcomes.
}
\label{fig::1}
\end{figure}

Here, we propose a postselection-free Lindbladian simulation method. It avoids the challenge of vanishing success probability using only one ancilla qubit. Our approach provides a translation of the classically known quantum jump algorithm~\cite{Dalibard1992,Dum1992,Klaus1993,Plenio1998,Daley2014} into executable quantum circuits. The key insight, as shown in Fig.~\ref{fig::1}, is to exactly match the stochasticity in quantum trajectories with the probabilistic outcomes of measurements in quantum circuits: each measurement result directly maps to a specific mathematical process in the quantum trajectory method, representing either a quantum jump or a non-unitary evolution step. In contrast to methods such as SND, LCU, and SVD, where certain measurement outcomes must be discarded, our protocol ensures that every measurement outcome is meaningful and contributes to the simulation.

Moreover, our scheme can be extended to probabilistically simulate a class of nonlinear Lindblad master equations (NLME)~\cite{Hugo2025,Lewalle2023,MCoppola2024,APaviglianiti2025} using two ancilla qubits. The NLME continuously interpolates between the LME and the evolution equation of effective non-Hermitian Hamiltonian (ENHH). It arises in monitored open quantum systems where quantum jumps are lost or postselected. Such equations provide a theoretical framework for probing the interplay between non-Hermitian effects and dissipation~\cite{Hugo2025}, as well as measurement-induced phenomena in imperfect settings~\cite{MCoppola2024,APaviglianiti2025}, processes of information loss~\cite{APaviglianiti2025} and quantum optimal control~\cite{Lewalle2023}. Conventional experimental proposals for realizing NLMEs rely on imperfect measurement processes~\cite{Hugo2025,Minganti2020}, which often encounter practical limitations due to uncontrollable efficiency of monitors. In contrast, our digital simulation scheme provides a flexible and highly controllable alternative.

%%%%%%%%%%%%%%%%%%%~~~~~~~~~~~%%%%%%%%%%%%%%%%%%%
%%%%%%%%%%%%%%%%%%%~~~~~~~~~~~%%%%%%%%%%%%%%%%

In this paper, we first present a trajectory-averaged interpretation of the NLME and propose a probabilistic 2-dilation implementation, which reduces to a 1-dilation for special cases (LME and ENHH). For LME, the scheme becomes fully deterministic due to absent post-selection effects. We analyze error scaling and validate the approach through dissipative XXZ model simulations. We further perform exploratory simulations of novel models featuring many-body localization in open systems and the postselected skin effect to demonstrate quantum computing's potential in investigating emerging phenomena in the Supplementary Material~\cite{SuMa}.

%%%%%%%%%%%%%%%%%%%~~~~~~~~~~~%%%%%%%%%%%%%%%%%%%
% \begin{figure*}[htbp]\centering
\begin{figure*}[htbp]\centering
\includegraphics[width=15cm]{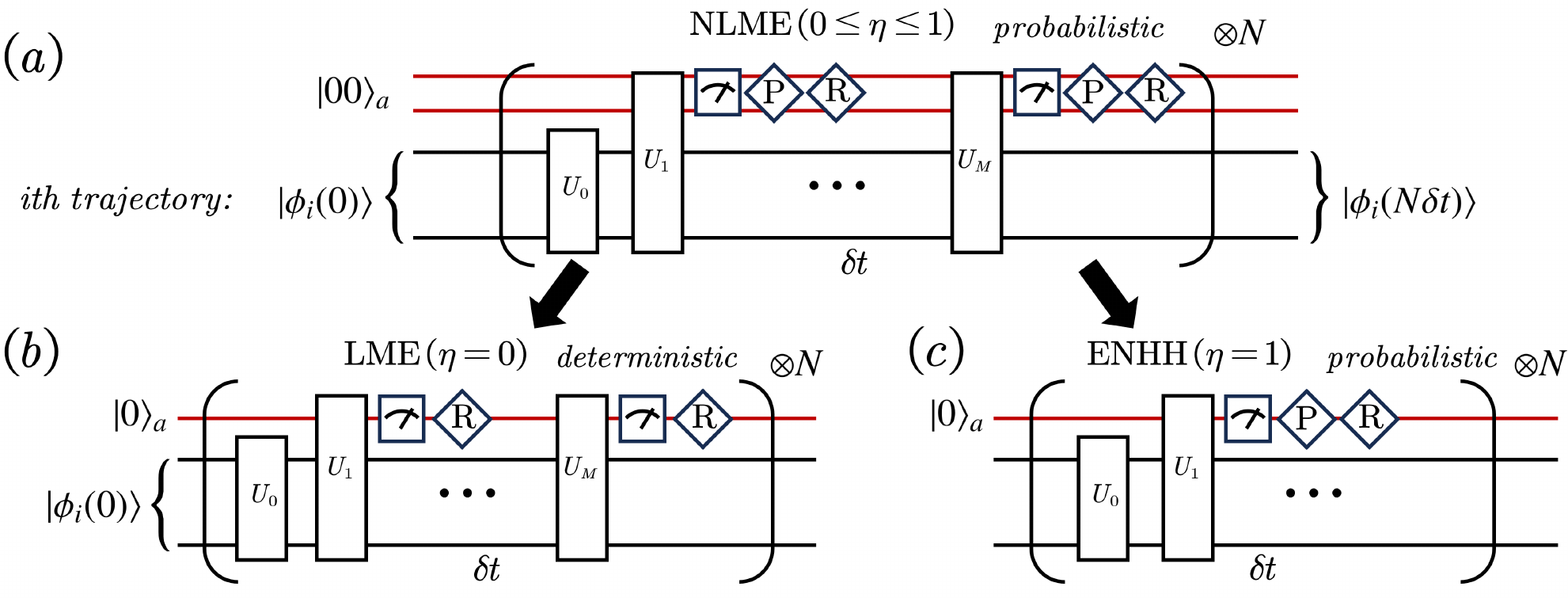}
\caption{ Scheme of the digital quantum trajectory simulation: (a) the 2-dilation method for the NLME as Eq.~(\ref{NLME}), (b) the 1-dilation method for the LME as Eq.~(\ref{LME}), and (c) the 1-dilation method for the evolution of ENHH as Eq.~(\ref{ENHH}). The system is initialized in state $|\phi(0)\rangle$. The auxiliary qubits are initialized in state $|00\rangle_a$ in (a) and in state $|0\rangle_a$ in both (b) and (c). At each time step $\delta t$, the operations inside the brackets are executed. The $U_0$ represents the Hamiltonian simulation of the time evolution $e^{-iH\delta t}$. The gate $U_{\mu \ge 1}$, corresponding to the $\mu$-th dissipation source, is given by Eq.~(\ref{U}) in (a), Eq.~(\ref{U0}) in (b), and Eq.~(\ref{U1}) in (c). After each measurement of auxiliary qubit, the postselection, denoted by 'P', is required only in (a) and (c), making their realization probabilistic. In contrast, the method in (b) is deterministic as it does not require postselection. In all three cases, the auxiliary qubit must be repeatedly reset to its initial state, denoted by 'R'.
}
\label{fig::2}
\end{figure*}
%%%%%%%%%%%%%%%%%%%~~~~~~~~~~~%%%%%%%%%%%%%%%%

%%%%%%%%%%%%%%%%%%%%%%%%%%%%%%%%%%%%%%%%%%%%%%%%%%%%%%%%%%%%%%%%%%%%%%%%%%%%%%%%%%%%%%%%%%%%%%%%%%%%%%%%%%%%%%%%%%%%%%%%
%%%%%%%%%%%%%%%%%%%%%%%%%%%%%%%%%%%%%%%%%%%%%%%%%%%%%%%%%%%%%%%%%%%%%%%%%%%%%%%%%%%%%%%%%%%%%%%%%%%%%%%%%%%%%%%%%%%%%%%%
\emph{ \color{blue} Trajectory-averaged interpretation of NLME.---}
The NLME is a generation of the LME~\cite{Hugo2025,Lewalle2023}. It continuously interpolates between LME and ENHH by reducing quantum jumps, given by
%%%%%%%%%%%%%%%%%%%%%%%%
\begin{equation}\label{NLME}
\begin{split}
& \frac{d}{dt} \rho = \mathcal{L}_{\eta}(\rho):= -i[H,\rho] + \sum_{\mu=1}^M \mathcal{D}_{\mu}(\rho),\\
& \mathcal{D}_{\mu}(\rho)= \gamma_\mu \left( -\frac{1}{2} \{ L_{\mu}^{\dag} L_{\mu}, \rho \} + (1-\eta_\mu) L_{\mu} \rho L_{\mu}^{\dag} + \eta_\mu  \langle L_{\mu}^{\dag} L_{\mu} \rangle \rho \right),
\end{split}
\end{equation}
%%%%%%%%%%%%%%%%%%%%%%%%
where the system with Hamiltonian $H$ couples to the environment by $M$ jump operators $L_\mu$ with strength $\gamma_\mu$. The $\eta_\mu$ ($0 \le \eta_\mu \le 1$) is the postselection strength that calibrates the reduction ratio of quantum jumps. When $\eta_\mu=0$ for all $\mu$, the NLME reduces to the standard LME,
%%%%%%%%%%%%%%%%%%%%%%%%
\begin{equation}\label{LME}
\frac{d\rho}{dt} =\mathcal{L}(\rho):= -i[H,\rho] + \sum_{\mu=1}^M \gamma_{\mu} \left( L_{\mu} \rho L_{\mu}^{\dag} - \frac{1}{2} \{ L_{\mu}^{\dag} L_{\mu}, \rho \} \right).
\end{equation}
%%%%%%%%%%%%%%%%%%%%%%%%
Conversely, when $\eta_\mu=1$ for all $\mu$, the NLME simplifies to the evolution of ENHH,
%%%%%%%%%%%%%%%%%%%%%%%%
\begin{equation}\label{ENHH}
\frac{d}{dt} \rho = -iH_{\rm eff}\rho + i\rho H^\dag_{\rm eff}  + \sum_{\mu=1}^M \gamma_\mu  \langle L_{\mu}^{\dag} L_{\mu} \rangle \rho,
\end{equation}
%%%%%%%%%%%%%%%%%%%%%%%%
where the ENHH takes the form $H_{\rm eff}=H-\frac{i}{2} \sum_{\mu=1}^M \gamma_\mu L_{\mu}^{\dag} L_{\mu}.$

The NLME can be interpreted by the quantum trajectory method~\cite{Hugo2025}. The expectation value of an arbitrary observable $\hat{O}$ is given by the average over many trajectories: $\langle \hat{O} \rangle = \Tr (\hat{O} \rho) = \lim_{K \to \infty} \frac{1}{K} \sum_{i=1}^K \langle \phi_i (t) | \, \hat{O} \, | \phi_i (t) \rangle$.
% %%%%%%%%%%%%%%%%%%%%%%%%
% \begin{equation}
% \langle \hat{O} \rangle = \Tr (\hat{O} \rho) = \lim_{K \to \infty} \frac{1}{K} \sum_{i=1}^K \langle \phi_i (t) | \, \hat{O} \, | \phi_i (t) \rangle.
% \end{equation}
% %%%%%%%%%%%%%%%%%%%%%%%%
The quantum state at time $t+\delta t$ for the $i$-th trajectory is obtained from the state at time $t$ through a stochastic evolution, as shown in Fig.~\ref{fig::1}(a):
%%%%%%%%%%%%%%%%%%%%%%%%
\begin{equation}\label{tadt}
\begin{split}
& |\phi_i (t+\delta t)\rangle=e^{-iH\delta t} \prod_{\mu=1}^{M} \U_\mu (\delta t) \  |\phi_i (t)\rangle, \\
& \U_\mu (\delta t) =\left\{ \begin{matrix} (1-\eta_\mu) \delta p_\mu : & \N L_\mu \\ 1-(1-\eta_\mu) \delta p_\mu : & \N \exp (-\frac{1}{2}\gamma_\mu L^\dag_{\mu}L_{\mu}\delta t) \end{matrix}  \right. \ ,
\end{split}
\end{equation}
%%%%%%%%%%%%%%%%%%%%%%%%
where $\N$ denotes normalization process and $\delta p_\mu = \gamma_\mu \delta t \langle \phi_i (t)|L^\dag_{\mu}L_{\mu} |\phi_i (t) \rangle$. The dissipation effect $\U_\mu (\delta t)$ represents a stochastic process where the system either jumps to the state $\N L_\mu |\phi (t)\rangle$ with the probability of $(1-\eta_\mu) \delta p_\mu$ or evolves to the state $\N \exp(-\frac{1}{2}\gamma_\mu L^\dag_{\mu}L_{\mu}\delta t)|\phi (t)\rangle$.

%%%%%%%%%%%%%%%%%%%%%%%%%%%%%%%%%%%%%%%%%%%%%%%%%%%%%%%%%%%%%%%%%%%%%%%%%%%%%%%%%%%%%%%%%%%%%%%%%%%%%%%%%%%%%%%%%%%%%%%%
%%%%%%%%%%%%%%%%%%%%%%%%%%%%%%%%%%%%%%%%%%%%%%%%%%%%%%%%%%%%%%%%%%%%%%%%%%%%%%%%%%%%%%%%%%%%%%%%%%%%%%%%%%%%%%%%%%%%%%%%
\emph{ \color{blue} Digital quantum simulation.---}
%%%\emph{ \color{blue} Probabilistic implementation of NLME simulation.---}
The key challenge in implementing an iterative simulation for a single quantum trajectory of NLME from Eq.~(\ref{tadt}) lies in how to realize the stochastic process $\U_\mu (\delta t)$. To address this within a quantum circuit, we assume a unitary gate $U_\mu$ that acts on the Hilbert space of the system along with two auxiliary qubits as follows:
%%%%%%%%%%%%%%%%%%%%%%%%
\begin{equation}\label{Uphi}
U_\mu \, |00\rangle_a |\phi\rangle = |00\rangle_a \, C_\mu |\phi\rangle + |01\rangle_a \, B_\mu |\phi\rangle + |10\rangle_a \, A_\mu |\phi\rangle,
\end{equation}
%%%%%%%%%%%%%%%%%%%%%%%%
where the subscript $a$ represents auxiliary qubits, and the state $|11\rangle_a $ is forbidden by the structure of $U_\mu$. After measuring the auxiliary qubits, the system probabilistically collapses into the state of $\N X_\mu |\phi\rangle$ ($X$ = $A$, $B$, or $C$) with a probability of $\langle X_{\mu}^\dag X_\mu \rangle := \langle \phi | X_{\mu}^\dag X_\mu | \phi \rangle$.

To match the measurement outcomes of Eq.~(\ref{Uphi}) with the stochastic results of $\U_\mu (\delta t)$, we assume that $A_\mu \propto \exp(-\frac{1}{2}\gamma_\mu L^\dag_{\mu}L_{\mu}\delta t) $ and $B_\mu \propto L_\mu $. To precisely match their probabilities, we discard the $C_\mu |\phi\rangle$ state by postselecting the auxiliary qubits to not be in the $|00\rangle_a$ state, and require renormalized probability distribution to satisfy
%%%%%%%%%%%%%%%%%%%%%%%%
\begin{equation}\label{PC}
\frac{\langle B_{\mu}^\dag B_\mu \rangle}{\langle A_{\mu}^\dag A_\mu \rangle+\langle B_{\mu}^\dag B_\mu \rangle}=(1-\eta_\mu) \delta p_\mu.
\end{equation}
%%%%%%%%%%%%%%%%%%%%%%%%
By enforcing both the probability constraint Eq.~(\ref{PC}) and the unitarity condition $U_{\mu}^\dag U_{\mu}=\mathbb{I} $, we obtain a solution for $U_{\mu}$ as
%%%%%%%%%%%%%%%%%%%%%%%%
\begin{equation}\label{U}
U_\mu = \left(
\begin{matrix}
C_\mu & B_\mu &  \tilde{A}_\mu & 0 \\ B_\mu & -C_\mu & 0 & \tilde{A}_\mu \\
A_\mu & 0 & -C_{\mu}^\dag & -B_{\mu}^\dag \\ 0 & A_\mu & -B_{\mu}^\dag & C_{\mu}^\dag
\end{matrix} \right) ,
\end{equation}
%%%%%%%%%%%%%%%%%%%%%%%%
where
% $A_\mu = \sqrt{1-\gamma_\mu \delta t L_{\mu}^\dag L_\mu}$, $\tilde{A}_\mu = \sqrt{1-\gamma_\mu \delta t L_\mu L_{\mu}^\dag}$, $B_\mu = \sqrt{(1-\eta_\mu)\gamma_\mu \delta t} \, L_\mu$, and $C_\mu = \sqrt{\eta_\mu \gamma_\mu \delta t} \, L_\mu$.
%%%%%%%%%%%%%%%%%%%%%%%%
\begin{equation}\label{ABC}
\begin{split}
& A_\mu = \sqrt{1-\gamma_\mu \delta t L_{\mu}^\dag L_\mu} \, , \ \  \tilde{A}_\mu = \sqrt{1-\gamma_\mu \delta t L_\mu L_{\mu}^\dag} \, , \\
& B_\mu = \sqrt{(1-\eta_\mu)\gamma_\mu \delta t} \, L_\mu \, ,  \ \  C_\mu = \sqrt{\eta_\mu \gamma_\mu \delta t} \, L_\mu \, ,
\end{split}
\end{equation}
%%%%%%%%%%%%%%%%%%%%%%%%
and note that $\exp(-\frac{1}{2}\gamma_\mu L^\dag_{\mu}L_{\mu}\delta t) = A_\mu + \mathcal{O}(\delta t)$.

The 2-dilation method for realizing the full trajectory is presented in Fig.~\ref{fig::2}(a). The system and auxiliary qubits are initialized in the state $|\phi_i(0)\rangle$ and $|00\rangle_a$. Then, $N$ iterations are carried out for a time evolution of duration $T=N \delta t$. During each time step $\delta t$, the unitary evolution $e^{-iH\delta t}$ governed by the system Hamiltonian $H$ is encoded in the logic gate $U_0$, which is an implemention by the Hamiltonian simulation~\cite{Childs2018} (e.g., via Trotterization). Each dissipation effect $\U_\mu (\delta t)$  is implemented by measuring and post-selecting the auxiliary qubits after applying the 2-dilation logic gate $U_\mu$ in Eq.~(\ref{U}), where only the measurement outcomes of $|01\rangle_a$ and $|10\rangle_a$ are considered valid. If the outcome is $|00\rangle_a$ the iteration is ended, and this trajectory is discarded. Following postselection, the auxiliary qubits are reset to $|00\rangle_a$ state in preparation for the next gate $U_{\mu+1}$.

Since the postselection requires that the measurement outcome $|00\rangle_a$ cannot appear throughout the entire trajectory, the success probability for a single trajectory, with M dissipation sources and $N \delta t$ time duration, is
%%%%%%%%%%%%%%%%%%%%%%%%
\begin{equation}\label{Pss}
\begin{split}
P &= \prod_{t=\delta t}^{N\delta t} \prod_{\mu=1}^{M} \left( 1-\langle \phi(t) | C_{\mu}^\dag C_{\mu} \phi(t) |\rangle  \right)\\
  &= \prod_{t=\delta t}^{N\delta t} \prod_{\mu=1}^{M} \left( 1-\eta_\mu \gamma_\mu \delta t \langle \phi(t) | L_{\mu}^\dag L_{\mu} |\phi(t) \rangle  \right).
\end{split}
\end{equation}
%%%%%%%%%%%%%%%%%%%%%%%%
This implies that the success probability follows an approximately exponential decay as a function of $T\times M \times \eta $.
% $P$ is roughly estimated as $(1-\bar{\eta} \bar{\gamma} \langle L_{\mu}^\dag L_{\mu} \rangle \delta t)^{NM}$, where $\bar{\eta}=\frac{1}{M}\sum_{\mu=1}^{M} \eta_\mu$ and $\bar{\gamma}=\frac{1}{M}\sum_{\mu=1}^{M} \gamma_\mu$.

% %%%%%%%%%%%%%%%%%%%%%%%%%%%%%%%%%%%%%%%%%%%%%%%%%%%%%%%%%%%%%%%%%%%%%%%%%%%%%%%%%%%%%%%%%%%%%%%%%%%%%%%%%%%%%%%%%%%%%%%%
% %%%%%%%%%%%%%%%%%%%%%%%%%%%%%%%%%%%%%%%%%%%%%%%%%%%%%%%%%%%%%%%%%%%%%%%%%%%%%%%%%%%%%%%%%%%%%%%%%%%%%%%%%%%%%%%%%%%%%%%%
% \emph{ \color{blue} Deterministic implementation of LME simulation.---}

When $\eta_\mu =0$ for all $\mu$, the NLME reduces to the standard LME in Eq.~(\ref{LME}). At the same time, $C_\mu=0$ and postselection success probability reaches unity, i.e., $P=1$ in Eq.~(\ref{Pss}), which indicates that the 2-dilation probabilistic implementation of the NLME can reduce to the 1-dilation deterministic implementation of the LME, as shown in Fig.~\ref{fig::2}(b). Note that it does not require postselection because the stochasticity in $\U_\mu(t)$ is completely equivalent to the randomness of measurement outcomes. This means that after measuring the auxiliary qubit in the state
%$U_\mu \, |0\rangle_a |\phi\rangle = |0\rangle_a \, B_\mu |\phi\rangle + |1\rangle_a \, A_\mu |\phi\rangle$,
%%%%%%%%%%%%%%%%%%%%%%%%
\begin{equation}\label{Uphi0}
U_\mu \, |0\rangle_a |\phi\rangle = |0\rangle_a \, B_\mu |\phi\rangle + |1\rangle_a \, A_\mu |\phi\rangle ,
\end{equation}
%%%%%%%%%%%%%%%%%%%%%%%%
where $U_\mu$ reduces to
%%%%%%%%%%%%%%%%%%%%%%%%
\begin{equation}\label{U0}
U_\mu = \left(
\begin{matrix}
B_\mu & \tilde{A}_\mu \\
A_\mu & -B_{\mu}^\dag
\end{matrix} \right),
\end{equation}
%%%%%%%%%%%%%%%%%%%%%%%%
if the outcome is $|1\rangle_a$, the system has an evolvtion of $\N \exp(-\frac{1}{2}\gamma_\mu L^\dag_{\mu}L_{\mu}\delta t)$ by the $A_\mu$, whereas the outcome of $|0\rangle_a$ corresponds to a quantum jump on the trajectory by the $B_\mu$. The consistency of the probabilities between $\U_\mu$ and measurement outcomes is guaranteed by the constraint equation Eq.~(\ref{PC}).

When $\eta_\mu =1$ for all $\mu$, the 2-dilation method for the NLME can also reduce to the 1-dilation method for the ENHH in Eq.~(\ref{ENHH}) due to $B_\mu=0$, as shown in Fig.~\ref{fig::2}(c). The gate $U_\mu$ and its action reduce to
%%%%%%%%%%%%%%%%%%%%%%%%
\begin{equation}\label{U1}
U_\mu \, |0\rangle_a |\phi\rangle = |0\rangle_a \, C_\mu |\phi\rangle + |1\rangle_a \, A_\mu |\phi\rangle;
\ \ U_\mu = \left(
\begin{matrix}
C_\mu & \tilde{A}_\mu \\
A_\mu & -C_{\mu}^\dag
\end{matrix} \right).
\end{equation}
%%%%%%%%%%%%%%%%%%%%%%%%
The postselection discards trajectories where the auxiliary qubit collapses to $|0\rangle_a$ after measurement.

%%%%%%%%%%%%%%%%%%%%%%%%%%%%%%%%%%%%%%%%%%%%%%%%%%%%%%%%%%%%%%%%%%%%%%%%%%%%%%%%%%%%%%%%%%%%%%%%%%%%%%%%%%%%%%%%%%%%%%%%
%%%%%%%%%%%%%%%%%%%%%%%%%%%%%%%%%%%%%%%%%%%%%%%%%%%%%%%%%%%%%%%%%%%%%%%%%%%%%%%%%%%%%%%%%%%%%%%%%%%%%%%%%%%%%%%%%%%%%%%%
\emph{ Error analysis and algorithmic scaling.---}
The errors in our quantum simulation schemes primarily stem from two sources: the random sampling error introduced during trajectory averaging, and the systematic error arising from the non-commutativity of operators within the $\delta t$ interval. According to the central limit theorem, the sampling error decreases as $1/\sqrt{K}$, where $K$ is the number of trajectories. Therefore, for bounded observables, the sampling error can be suppressed below the level of the systematic error by increasing the number of trajectories.

For the systematic error, our algorithm maintains first-order accuracy within each $\delta t$. Consequently, for an NLME simulation of total time $t$ targeting a precision $ \varepsilon$, the required circuit depth scales as $\mathcal{O}(t^2/\varepsilon)$. For the LME simulation, further results can be obtained through diamond norm analysis (see Supplementary Material~\cite{SuMa}). Specifically, when $U_0$ is first-order accurate in $\delta t$, the scaling behavior of the circuit depth is
%%%%%%%%%%%%%%%%%%%%%%%%
\begin{equation}
\mathcal{O} \left(M \frac{\lambda^2 t^2}{\varepsilon}\right).
\end{equation}
%%%%%%%%%%%%%%%%%%%%%%%%
Here, $M$ is the number of dissipative channels, and $\lambda=\lVert H \rVert + M \gamma_x \lVert L_x \rVert^2$, where $\gamma_x \lVert L_x \rVert^2$ is the maximum value in the set of spectral norms: $\{ \gamma_1 \lVert L_1 \rVert^2,  \gamma_2 \lVert L_2 \rVert^2, \cdots \gamma_M \lVert L_M \rVert^2  \}$.

%%%%%%%%%%%%%%%%%%%%%%%%%%%%%%%%%%%%%%%%%%%%%%%%%%%%%%%%%%%%%%%%%%%%%%%%%%%%%%%%%%%%%%%%%%%%%%%%%%%%%%%%%%%%%%%%%%%%%%%%
%%%%%%%%%%%%%%%%%%%%%%%%%%%%%%%%%%%%%%%%%%%%%%%%%%%%%%%%%%%%%%%%%%%%%%%%%%%%%%%%%%%%%%%%%%%%%%%%%%%%%%%%%%%%%%%%%%%%%%%%
\emph{ Numerical experiments.---}
We numerically test our scheme by simulating different dynamical processes. A detailed pedagogical example of a two-level atom monitored through its spontaneous emission~\cite{Hugo2025} is provided in the Supplemental Material~\cite{SuMa}, to demonstrate the validity of our NLME simlation approach.

\begin{figure}[htbp]\centering
\includegraphics[width=8.5cm]{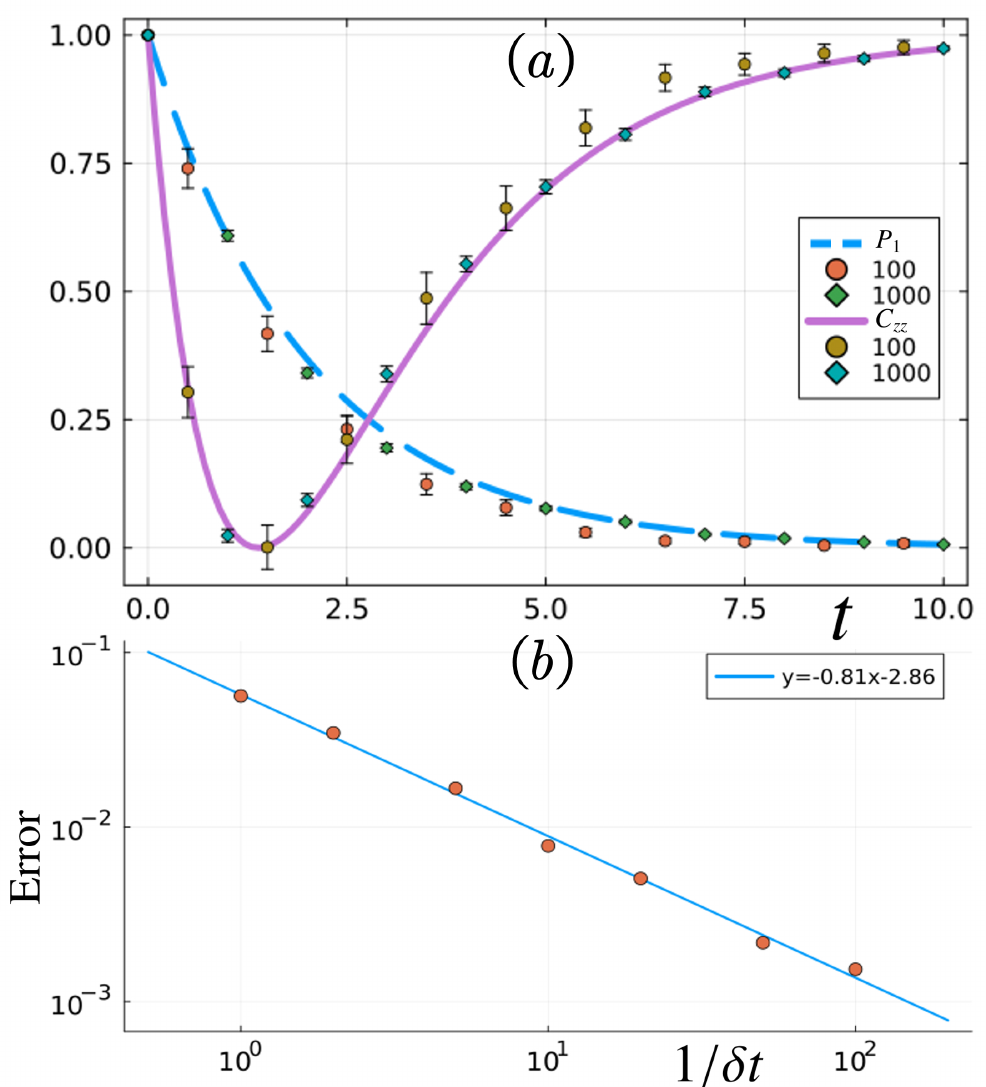}
\caption{ Numerical simulation of dissipative XXZ spin chain. We study the evolution of a 5-site dissipative XXZ chain from $t=1$ to $10$. The system is initialized in the all-spin-up state, with parameters $(J, \Delta , \gamma)=(1, 2, 0.5)$. (a) Exact results of the spin-up probability at the first site, $P_1= {\rm Tr} (\sigma_{1}^+ \sigma_{1}^- \rho)$, and the averaged nearest-neighbor correlations, $C_{zz}=\frac{1}{L-1}\sum_{j=1}^{L-1} \langle \sigma_{l+1}^z \sigma_l^z \rangle$, are shown as the blue dashed line and the solid magenta line, respectively. Circular and rhombus markers with error bars denote the results of our digital simulation method with time step $\delta t =0.1$, using 100 and 1000 trajectories, respectively. (b) Scaling of the error of the density matrix $\rho$ at $t=10$ with the simulation time step $\delta t$, obtained by averaging over 100,000 trajectories. Data ponts correspond to $\delta t = (1.0,0.5,0.2,0.1,0.05,0.02,0.01)$, and the solid line shows a fit to the data.
}
\label{fig::3}
\end{figure}

Here, we present results for a dissipative XXZ spin chain governed by the LME:
%%%%%%%%%%%%%%%%%%%%%%%%
\begin{equation}\label{LXXZ}
\frac{d}{dt} \rho = -i[H,\rho]+\gamma \sum_{l=1}^L \left( -\frac{1}{2} \{ \sigma_l^+ \sigma_l^-, \rho \} + \sigma_l^- \rho  \sigma_l^+ \right),
\end{equation}
%%%%%%%%%%%%%%%%%%%%%%%%
with the Hamiltonian
%%%%%%%%%%%%%%%%%%%%%%%%
\begin{equation}\label{HXXZ}
H=J\sum_{l=1}^{L-1} \left(  \sigma_{l+1}^x \sigma_l^x +  \sigma_{l+1}^y \sigma_l^y + \Delta \sigma_{l+1}^z \sigma_l^z  \right),
\end{equation}
%%%%%%%%%%%%%%%%%%%%%%%%
where $\sigma_l^{\pm}$ and $\sigma_l^{x,y,z}$ are the spin raising/lowering operators and Pauli matrices at site $l$, respectively. We compare the numerical simulation of the Eq.~(\ref{LXXZ}) presented in Fig.~\ref{fig::2}(b) with the exact solutions, as shown in Fig.~\ref{fig::3}, where the $U_0$ is implemented using a second-order Trotter-Suzuki decomposition for the Hamiltonian in Eq.~(\ref{HXXZ}) split into even and odd terms. As shown in Fig.~\ref{fig::3}(a), the numerical results for the spin-up probability at the first site ($\rm P_1$) and the averaged nearest-neighbor correlations ($C_{zz}=\frac{1}{L-1}\sum_{j=1}^{L-1} \langle \sigma_{l+1}^z \sigma_l^z \rangle$) agree well with the exact solutions. The associated standard error decreases significantly as time and the number of trajectories increase. Meanwhile, we study the dependence of the numerical error of the density matrix on the time step $\delta t$. Here, the error is defined as the trace norm of the difference between $\rho'$ obtained from our trajectory-based method and the exact $\rho$, i.e., $error=\lVert \rho'-\rho \rVert_1 $, evaluated at the final time $T=10$. As shown in Fig.~\ref{fig::3}(b), the error decreases as $\delta t$ becomes smaller, nearly scaling as $\delta t ^ {0.8}$.
%%%%%%%%%%%%%%%%%%%%%%%%%%%%%%%%%%%%%%%%%%%%%%%%%
%%%%%%%%%%%%%%%%%%%~~~~~~~~~~~%%%%%%%%%%%%%%%%%%%
%%%%%%%%%%%%%%%%%%%~~~~~~~~~~~%%%%%%%%%%%%%%%%
%%%%%%%%%%%%%%%%%%%%%%%%%%%%%%%%%%%%%%%%%%%%%%%%% 

Furthermore, we explore two complicated phenomena: the postselected skin effect~\cite{Hugo2025} and open-system localization~\cite{Yusipov2017,Vakulchyk2018,YLiu2024} in the Supplementary Material~\cite{SuMa}. These remain challenging to address with current analog simulations or classical computation. The latter, in particular, involving dissipation-induced interactions, could be considered as an open question of many-body localization in open system. The preliminary small-scale simulation results are provided to imply the potential advantage of quantum computing in exploring novel problems.

%%%%%%%%%%%%%%%%%%%%%%%%%%%%%%%%%%%%%%%%%%%%%%%%%%%%%%%%%%%%%%%%%%%%%%%%%%%%%%%%%%%%%%%%%%%%%%%%%%%%%%%%%%%%%%%%%%%%%%%%
%%%%%%%%%%%%%%%%%%%%%%%%%%%%%%%%%%%%%%%%%%%%%%%%%%%%%%%%%%%%%%%%%%%%%%%%%%%%%%%%%%%%%%%%%%%%%%%%%%%%%%%%%%%%%%%%%%%%%%%%
\emph{\color{blue} Summary.---}
We tackle the scalability challenge in numerically simulating the LME over long-time evolutions with multiple dissipation sources. Based on quantum trajectory averaging, we propose a digital quantum simulation scheme for dynamics governed by the LME, ENHH, and NLME. The simulation of the NLME is achieved via a 2-dilation method, which simplifies to 1-dilation for the LME and ENHH. Notably, our approach enables deterministic simulation of the standard LME without postselection, distinguishing it from existing probabilistic schemes and allowing for long-time simulations of complex open systems.

Our work paves the way for efficient digital quantum simulations of the LME, enables the exploration of the interplay between non-Hermitian Hamiltonians and pure dissipation processes through NLME simulations, and offers a promising simulation scheme for various cutting-edge theoretical models in open quantum systems and non-Hermitian physics.

%%%%%%%%%%%%%%%%%%%%%%%%%%%%%%%%%%%%%%%%%%%%%%%%%%%%%%%%%%%%%%%%%%%%%%%%%%%%%%%%%%%%%%%%%%%%%%%%%%%%%%%%%%%%%%%%%%%%%%%%
%%%%%%%%%%%%%%%%%%%%%%%%%%%%%%%%%%%%%%%%%%%%%%%%%%%%%%%%%%%%%%%%%%%%%%%%%%%%%%%%%%%%%%%%%%%%%%%%%%%%%%%%%%%%%%%%%%%%%%%%
\begin{acknowledgments}
\emph{Acknowledgments.---}
Y.-G. Liu thanks Zongsheng Zhou, Jiaheng Li, Sirui Peng, Hongyi Zhou, Jinzhao Sun and Xiao Yuan  for the valuable discussions. The work is supported by National Key Research and Development Program of China (Grant Nos. 2021YFA1402104 and 2023YFA1406704),  the National Natural Science Foundation of China (Grants Nos. 12474287, 12174436, T2121001, 92265207).
\end{acknowledgments}

%%%%%%%%%%%%%%%%%%%%%%%%%%%%%%%%%%%%%%%%%%%%%%%%%%%%%%%%%%%%%%%%%%%%%%%%%%%%%%%%%%%%%%%%%%%%%%%%%%%%%%%%%%%%%%%%%%%%%%%%
%%%%%%%%%%%%%%%%%%%%%%%%%%%%%%%%%%%%%%%%%%%%%%%%%%%%%%%%%%%%%%%%%%%%%%%%%%%%%%%%%%%%%%%%%%%%%%%%%%%%%%%%%%%%%%%%%%%%%%%%

\clearpage
\widetext

\setcounter{equation}{0}
\def\theequation{S\arabic{equation}}
\setcounter{figure}{0}
\def\thefigure{S\arabic{figure}}
%\appendix

\section{Supplemental Material: Digital Quantum Simulation of the Nonlinear Lindblad Master Equation Based on Quantum Trajectory Averaging}

\begin{itemize}
    \item S1. \ Systematic error of our LME simulation
    \item S2.\ Two-level atom under monitoring spontaneous emission
      \begin{itemize}
      \item Exact solution via vectorization
      \item Digital simulation and error analysis
      \end{itemize}
    \item S3.\ Simulation of many-body localization in open system and the postselected skin effect
      \begin{itemize}
      \item Many-body localization in open system
      \item Postselected skin effect
      \end{itemize}
\end{itemize}
%%%%%%%%%%%%%%%%%%%%%%%%%%%%%%%%%%%%%%%%%%%%%%%%%%%%%%%%%%%%%%%%%%%%%%%%%%%%%%%%%%%%%%%%%%%%%%%%%%%%%%%%%%%%%%%%%%%%%%%%
%%%%%%%%%%%%%%%%%%%%%%%%%%%%%%%%%%%%%%%%%%%%%%%%%%%%%%%%%%%%%%%%%%%%%%%%%%%%%%%%%%%%%%%%%%%%%%%%%%%%%%%%%%%%%%%%%%%%%%%%
\subsection{S1.\ Systematic error of our LME simulation}
The dynamics of the Lindblad master equation (LME) is governed by
%%%%%%%%%%%%%%%%%%%%%%%%
\begin{equation}
\frac{d\rho}{dt} =\mathcal{L}(\rho):= -i[H,\rho] + \sum_{\mu=1}^M \gamma_{\mu} \left( L_{\mu} \rho L_{\mu}^{\dag} - \frac{1}{2} \{ L_{\mu}^{\dag} L_{\mu}, \rho \} \right).
\end{equation}
%%%%%%%%%%%%%%%%%%%%%%%%
Our trajectory-based simulation scheme of the above equation, as shown in FIG.2 (b) in the main text, generates an ensemble of pure states. Within each time step $\delta t$, this stochastic process is equivalently described by a quantum map $\mathcal{E}_{\delta t}$:
%%%%%%%%%%%%%%%%%%%%%%%%
\begin{equation} \label{QM}
\mathcal{E}_{\delta t}(\rho)= U_0 \, \prod_{\mu=1}^{M} \mathcal{E}_\mu (\rho) \, U_0^\dag,
\end{equation}
%%%%%%%%%%%%%%%%%%%%%%%%
where $U_0$ is a circuit implementation of unitary evolution $e^{-iH\delta t}$, and
%%%%%%%%%%%%%%%%%%%%%%%%
\begin{equation}
\mathcal{E}_\mu (\rho) = A_\mu \rho A_\mu^\dag + B_\mu \rho B_\mu^\dag,
\end{equation}
%%%%%%%%%%%%%%%%%%%%%%%%
where $A_\mu = \sqrt{1-\gamma_\mu \delta t L_\mu^\dag L_\mu}$ and $B_\mu = \sqrt{\gamma_\mu \delta t} L_\mu$. 

The error of Eq.~(\ref{QM}) relative to the standard evolution of the LME can be described by the diamond norm:
%%%%%%%%%%%%%%%%%%%%%%%%
\begin{equation}
\lVert \mathcal{E}_{\delta t} - e^{\mathcal{L} \delta t} \rVert_\diamond := \sup\limits_{Q \in \mathcal{H} \otimes \mathcal{K}}  \Big\lVert [(\mathcal{E}_{\delta t} - e^{\mathcal{L} \delta t} ) \otimes I_{\mathcal{K}}] (Q) \Big\rVert_1,
\end{equation}
%%%%%%%%%%%%%%%%%%%%%%%%
where $\lVert \cdot \rVert_1$ is the trace norm, $I_{\mathcal{K}}$ is the identity matrix on the Hilbert space $\mathcal{K}$, which has the same dimension as the Hilbert space $\mathcal{H}$ of the Hamiltonian. $Q$ is a density matrix in space $\mathcal{H} \otimes \mathcal{K}$.

By the triangle inequality, the upper bound of error can be evaluated as
%%%%%%%%%%%%%%%%%%%%%%%%
\begin{equation} \label{BofE}
\lVert \mathcal{E}_{\delta t} - e^{\mathcal{L} \delta t} \rVert_\diamond  \le \lVert \mathcal{E}_{\delta t} -\mathbf{1} - \delta t \mathcal{L}  \rVert_\diamond  + \lVert \mathbf{1} + \delta t \mathcal{L} -  e^{\mathcal{L} \delta t} \rVert_\diamond.
\end{equation}
%%%%%%%%%%%%%%%%%%%%%%%%
The second term is bounded by the result of Cleve et al.  ~\cite{RCleve2017}:
%%%%%%%%%%%%%%%%%%%%%%%%
\begin{equation} \label{Sec}
\lVert \mathbf{1} + \delta t \mathcal{L} -  e^{\mathcal{L} \delta t} \rVert_\diamond \le  ( \delta t \lVert \mathcal{L} \rVert_\diamond )^2 \le 4 \lambda^2 \delta t^2,  
\end{equation}
%%%%%%%%%%%%%%%%%%%%%%%%
where $\lVert \cdot \rVert$ (without a subscript) denotes the spectral norm, and the parameter $\lambda$ is defined as
%%%%%%%%%%%%%%%%%%%%%%%%
\begin{equation}
\lambda =  \lVert H \rVert + M \gamma_x \lVert L_x \rVert^2,
\end{equation}
%%%%%%%%%%%%%%%%%%%%%%%%
with 
%%%%%%%%%%%%%%%%%%%%%%%%
\begin{equation} \label{rxLx}
\gamma_x \lVert L_x \rVert^2 = \max \Big\lbrace \gamma_1 \lVert L_1 \rVert^2,  \gamma_2 \lVert L_2 \rVert^2, \cdots , \gamma_M \lVert L_M \rVert^2  \Big\rbrace.
\end{equation}
%%%%%%%%%%%%%%%%%%%%%%%%
The second inequality in Eq.~(\ref{Sec}) follows from the norm inequality
%%%%%%%%%%%%%%%%%%%%%%%%
\begin{equation}~\label{IEN}
\|A Q B\|_{1} \leq \|A\| \, \|B\|  \,  \|Q\|_{1} =  \|A\| \, \|B\|  
\end{equation}
%%%%%%%%%%%%%%%%%%%%%%%%

The error of the first term, 
%%%%%%%%%%%%%%%%%%%%%%%%
\begin{equation}
\| \mathcal{E}_{\delta t} -\mathbf{1} - \delta t \mathcal{L} \|_\diamond = \sup\limits_{Q \in \mathcal{H} \otimes \mathcal{K}}  \Big\| [(\mathcal{E}_{\delta t} -\mathbf{1} - \delta t \mathcal{L}) \otimes I_{\mathcal{K}}] (Q) \Big\|_1,
\end{equation}
%%%%%%%%%%%%%%%%%%%%%%%%
depends on the Hamiltonian simulation $U_0$. Here, we consider the case where $U_0$ is implemented with a first-order accuracy in $\delta t$. To evaluate this error, we we retain terms up to second order in $\delta t$:
%%%%%%%%%%%%%%%%%%%%%%%%
\begin{subequations}
\begin{align}
& A_\mu=\sqrt{1-\gamma_\mu \delta t L^\dag_\mu L_\mu} = 1-\frac{1}{2}\gamma_\mu \delta t L^\dag_\mu L_\mu -\frac{1}{8}\gamma_\mu^2 \delta t^2 (L^\dag_\mu L_\mu)^2 + \mathcal{O} (\delta t^3) \\
& U_0 = 1-iH\delta t -\frac{1}{2} \delta t^2 H^2 + \mathcal{O} (\delta t^3).
\end{align}
\end{subequations}
%%%%%%%%%%%%%%%%%%%%%%%%
Then we have
%%%%%%%%%%%%%%%%%%%%%%%%
\begin{equation}
\begin{split}
& \| [(\mathcal{E}_{\delta t} -\mathbf{1} - \delta t \mathcal{L}) \otimes I_{\mathcal{K}}] (Q)\|_1 \\
& = \delta t^2 \left( \Big\| HQH -\frac{1}{2} \{ H^2, Q \} - i [H , \sum_\mu d_\mu (Q)] + \sum_\mu j_\mu (Q) + \sum_{m>n} d_m d_n (Q) \Big\|_1 \right) + \mathcal{O}(\delta t^3) \\
& \le \delta t^2 \left( 2\| H \|^2 + 2 \| H \| \, \Big\| \sum_\mu d_\mu (Q)\Big\|_1  +  \Big\| \sum_\mu j_\mu (Q) \Big\|_1 +  \Big\| \sum_{m>n} d_m d_n (Q) \Big\|_1 \right) + \mathcal{O}(\delta t^3), \\
\end{split}
\end{equation}
%%%%%%%%%%%%%%%%%%%%%%%%
where the superoperators $d_\mu$ and $j_\mu$ are defined as
%%%%%%%%%%%%%%%%%%%%%%%%
\begin{subequations}
\begin{align}
d_\mu (\rho) &= \gamma_\mu \left(  L_\mu \rho L_\mu^\dag - \frac{1}{2} \{L_\mu^\dag L_\mu , \rho   \}    \right), \\
j_\mu (\rho) &= \frac{1}{4} \gamma_\mu^2 \left(  L_\mu^\dag L_\mu \rho  L_\mu^\dag L_\mu -\frac{1}{2} \{ (L_\mu^\dag L_\mu)^2 , \rho \}  \right).
\end{align}
\end{subequations}
%%%%%%%%%%%%%%%%%%%%%%%%
By Eq.~(\ref{rxLx}) and Eq.~(\ref{IEN}), we obtain the following bounds:
%%%%%%%%%%%%%%%%%%%%%%%%
\begin{subequations}
\begin{align}
& \Big\| \sum_\mu d_\mu (Q) \Big\|_1   \le \sum_{\mu} \gamma_\mu \| L_\mu \|^2 \le M \gamma_x \| L_x \|^2, \\
& \Big\| \sum_\mu j_\mu (Q) \Big\|_1  \le \sum_\mu \frac{1}{2} \gamma_\mu^2 \| L_\mu^\dag L_\mu \|^2 \le \frac{M}{2} \gamma_x^2 \| L_x \|^4. \\
& \Big\| \sum_{m>n} d_m d_n (Q) \Big\|_1  \le 2 M (M-1) \gamma_x^2 \| L_x \|^4,
\end{align}
\end{subequations}
%%%%%%%%%%%%%%%%%%%%%%%%
Furthermore,
%%%%%%%%%%%%%%%%%%%%%%%%
\begin{equation}
\begin{split}
& \| [(\mathcal{E}_{\delta t} -\mathbf{1} - \delta t \mathcal{L}) \otimes I_{\mathcal{K}}] (Q)\|_1 \\
& \le \delta t^2 \left( 2 \| H \|^2 + 2M \| H \| \gamma_x \| L_x \|^2 + (2M^2 -1.5M) \gamma_x^2 \| L_x \|^4 \right) + \mathcal{O}(\delta t^3), \\ 
& \le \delta t^2 2 ( \| H \| + M \gamma_x \| L_x \|^2)^2.
\end{split}
\end{equation}
%%%%%%%%%%%%%%%%%%%%%%%%
Combining these results, we finally obtain
%%%%%%%%%%%%%%%%%%%%%%%%
\begin{equation} 
\| \mathcal{E}_{\delta t} -\mathbf{1} - \delta t \mathcal{L}  \|_\diamond \le 2\lambda^2 \delta t^2,
\end{equation}
%%%%%%%%%%%%%%%%%%%%%%%%
and
%%%%%%%%%%%%%%%%%%%%%%%%
\begin{equation} \label{Edt}
\lVert \mathcal{E}_{\delta t} - e^{\mathcal{L} \delta t} \rVert_\diamond  \le  6\lambda^2 \delta t^2.
\end{equation}
%%%%%%%%%%%%%%%%%%%%%%%%

Now consider simulating the dynamics for a total time $t = N \delta t$ with a target precision $\varepsilon$. Eq.~(\ref{Edt}) implies that the error per time step could satisfy $6\lambda^2 \delta t^2 \leq \varepsilon / N$. Consequently, the required number of time steps $N$ scales as $\mathcal{O} (\lambda^2 t^2 / \varepsilon ) $ and the overall circuit depth scales as $\mathcal{O} ( M \lambda^2 t^2 / \varepsilon ) $.

%%%%%%%%%%%%%%%%%%%%%%%%%%%%%%%%%%%%%%%%%%%%%%%%%%%%%%%%%%%%%%%%%%%%%%%%%%%%%%%%%%%%%%%%%%%%%%%%%%%%%%%%%%%%%%%%%%%%%%%%
%%%%%%%%%%%%%%%%%%%%%%%%%%%%%%%%%%%%%%%%%%%%%%%%%%%%%%%%%%%%%%%%%%%%%%%%%%%%%%%%%%%%%%%%%%%%%%%%%%%%%%%%%%%%%%%%%%%%%%%%
\subsection{S2.\ Two-level atom under monitoring spontaneous emission}

As a pedagogical example to introduce our NLME simulation algorithm and demonstrate its validity, we consider the simulation of a two-level atom under monitoring of its spontaneous emission. Its dynamics is decribed by the LME,
%%%%%%%%%%%%%%%%%%%%%%%%
\begin{equation}
\frac{d}{dt} \rho = -i[H,\rho] + \gamma \left( -\frac{1}{2} \{ L^\dag L, \rho \} + L \rho L^\dag \right),
\end{equation}
%%%%%%%%%%%%%%%%%%%%%%%%
where the atom is in the space of ground state $|g\rangle$ and excited state $|e\rangle$, $H=J\sigma^x$ is the driven Hamiltonian, $L=|g\rangle\langle e|=\sigma^-$ is the dissipative operator and $\gamma$ is the spontaneous emission rate. The spontaneously emitted photons are monitored by an $\eta$-efficiency detector. In the postselection experiment, the observed data of atomic population is discarded when emission photons are detected simultaneously. With the remaining data, the evolution of atomic system is rebuilt by the nonlinear Lindblad master equation (NLME):
%%%%%%%%%%%%%%%%%%%%%%%%
\begin{equation}~\label{AtomE}
\frac{d}{dt} \rho = -iJ[\sigma^x,\rho] + \gamma \left( -\frac{1}{2} \{ \sigma^+ \sigma^-, \rho \} + (1-\eta) \sigma^- \rho \sigma^+ + \eta  \langle \sigma^+ \sigma^- \rangle \rho \right).
\end{equation}
%%%%%%%%%%%%%%%%%%%%%%%%

%%%%%%%%%%%%%%%%%%%%%%%%%%%%%%%%%%%%%%%%%%%%%%%%%%%%%%%%%%%%%%%%%%%%%%%%%%%%%%%%%%%%%%%%%%%%%%%%%%%%%%%%%%%%%%%%%%%%%%%%
\subsubsection{Exact solution via vectorization}

To benchmark our digital simulation, we require the exact solution of Eq.~(\ref{AtomE}) via the vectorization method. In this method, the density matrix $\rho$ is mapped to the density vector $|\rho\rangle$:
%%%%%%%%%%%%%%%%%%%%%%%%
\begin{equation}
\rho= \left(\begin{matrix}  \rho_{ee} & \rho_{eg} \\ \rho_{ge} & \rho_{gg} \end{matrix} \right) \ \ \to \ \ |\rho\rangle= (\rho_{ee}, \  \rho_{eg}, \ \rho_{ge}, \  \rho_{gg})^{\RT},
\end{equation}
%%%%%%%%%%%%%%%%%%%%%%%%
and the NLME is mapped as
%%%%%%%%%%%%%%%%%%%%%%%%
\begin{equation}
\frac{d}{dt}\,|\rho\rangle\,=\,\RL_\eta \, |\rho\rangle,
\end{equation}
%%%%%%%%%%%%%%%%%%%%%%%%
where
%%%%%%%%%%%%%%%%%%%%%%%%
\begin{equation}
\RL_\eta=-i\RH \otimes \RI + i \RI \otimes \RH^{\RT} -\frac{\gamma}{2} (\RL^\dag \RL \otimes \RI + \RI \otimes \RL^{\RT} \RL^* ) + (1-\eta)\gamma \RL \otimes \RL^* + \eta \gamma <\RL^\dag \RL>,
\end{equation}
%%%%%%%%%%%%%%%%%%%%%%%%
where
%%%%%%%%%%%%%%%%%%%%%%%%
\begin{equation} \label{NLMEatom}
\RH=\left( \begin{matrix} 0 & J \\ J & 0  \end{matrix} \right), \ \ \RL=\left( \begin{matrix} 0 & 0 \\ 1 & 0  \end{matrix} \right), \ \ \RI=\left( \begin{matrix} 1 & 0 \\ 0 & 1  \end{matrix} \right).
\end{equation}
%%%%%%%%%%%%%%%%%%%%%%%%
Then we get the vectorized NLME:
%%%%%%%%%%%%%%%%%%%%%%%%
\begin{equation}
\frac{d}{dt}\left(\begin{matrix} \rho_{ee} \\ \rho_{eg} \\ \rho_{ge} \\ \rho_{gg} \end{matrix}  \right) =
\left(\begin{matrix}
-iJ(\rho_{ge}-\rho_{eg}) +\eta \gamma  \rho_{ee}^2 - \gamma \rho_{ee} \\
-iJ(\rho_{gg}-\rho_{ee}) + \eta \gamma \rho_{eg} \rho_{ee} -0.5\gamma\rho_{eg} \\
iJ(\rho_{gg}-\rho_{ee}) + \eta \gamma \rho_{ge} \rho_{ee} -0.5\gamma\rho_{ge} \\
iJ(\rho_{ge}-\rho_{eg}) + \eta \gamma \rho_{gg} \rho_{ee} + (1-\eta)\gamma\rho_{ee}.
\end{matrix}  \right),
\end{equation}
%%%%%%%%%%%%%%%%%%%%%%%%
which can be solved by the Runge-Kutta methods.

%%%%%%%%%%%%%%%%%%%%%%%%%%%%%%%%%%%%%%%%%%%%%%%%%%%%%%%%%%%%%%%%%%%%%%%%%%%%%%%%%%%%%%%%%%%%%%%%%%%%%%%%%%%%%%%%%%%%%%%%
\subsubsection{Digital simulation and error analysis }

Our scheme of digital simulation of Eq.~(\ref{AtomE}) is illustrated in Fig.2 in the main text. The 2-dilation method (Fig.2 (a)) is applicable for all values of $\eta$. Specifically, for $\eta=0$ and $\eta=1$, the simulation simplifies to the 1-dilation method, as shown in Fig.2 (b) and (c). We take the 2-dilation method as an example. We assume that the atom starts at the excited state, which is represented as the system qubit at the state $|\phi(0)\rangle=(1,0)^{\RT}$. For each time step $\delta t$ in a single trajectory simulation, we first apply the gate $U_0=e^{-iJ\sigma^x \delta t}$ on $|\phi(t)\rangle$, followed by the gate $U_1$ on the state $|00\rangle_a \otimes U_0|\phi(t)\rangle$, where $|00\rangle_a=(1,0,0,0)^{\RT}$ is a state of two auxiliary qubits and $U_1$ is given by Eq.~(7) in the main text, with
%%%%%%%%%%%%%%%%%%%%%%%%
\begin{equation}\label{ABC1}
\begin{split}
& A_1 = \sqrt{1-\gamma \delta t L^\dag L}= \left( \begin{matrix} \sqrt{1-\gamma \delta t} & 0 \\ 0 & 1   \end{matrix}\right) \\
& \tilde{A}_1 = \sqrt{1-\gamma \delta t L L^\dag}= \left( \begin{matrix}  1 & 0 \\ 0 & \sqrt{1-\gamma \delta t}  \end{matrix}\right) \\
& B_1 = \sqrt{(1-\eta)\gamma \delta t} \, L = \left( \begin{matrix}  0 & 0 \\ \sqrt{(1-\eta)\gamma\delta t} & 0  \end{matrix}\right) \\
& C_1 = \sqrt{\eta \gamma \delta t} \, L= \left( \begin{matrix}  0 & 0 \\ \sqrt{\eta \gamma\delta t} & 0  \end{matrix}\right) \, .
\end{split}
\end{equation}
%%%%%%%%%%%%%%%%%%%%%%%%
We measure the auxiliary qubits and reset them to the state $|00\rangle_a$. Repeating the above steps N times results the state at the target time $|\phi(N\delta t)\rangle$. The postselection requires that in the N measurements, no outcomes should be $|00\rangle_a$; otherwise, this trajectory will be discarded. After multiple experimental rounds, a sufficient number of valid trajectories are collected. Averaging over these trajectories yields the desired simulation results. Let $K$ denote the total number of experimental rounds and $K_{eff}$ the number of valid trajectories obtained. 

%%%%%%%%%%%%%%%%%%%~~~~~~~~~~~%%%%%%%%%%%%%%%%%%%
\begin{figure*}[htbp]\centering
\includegraphics[width=14cm]{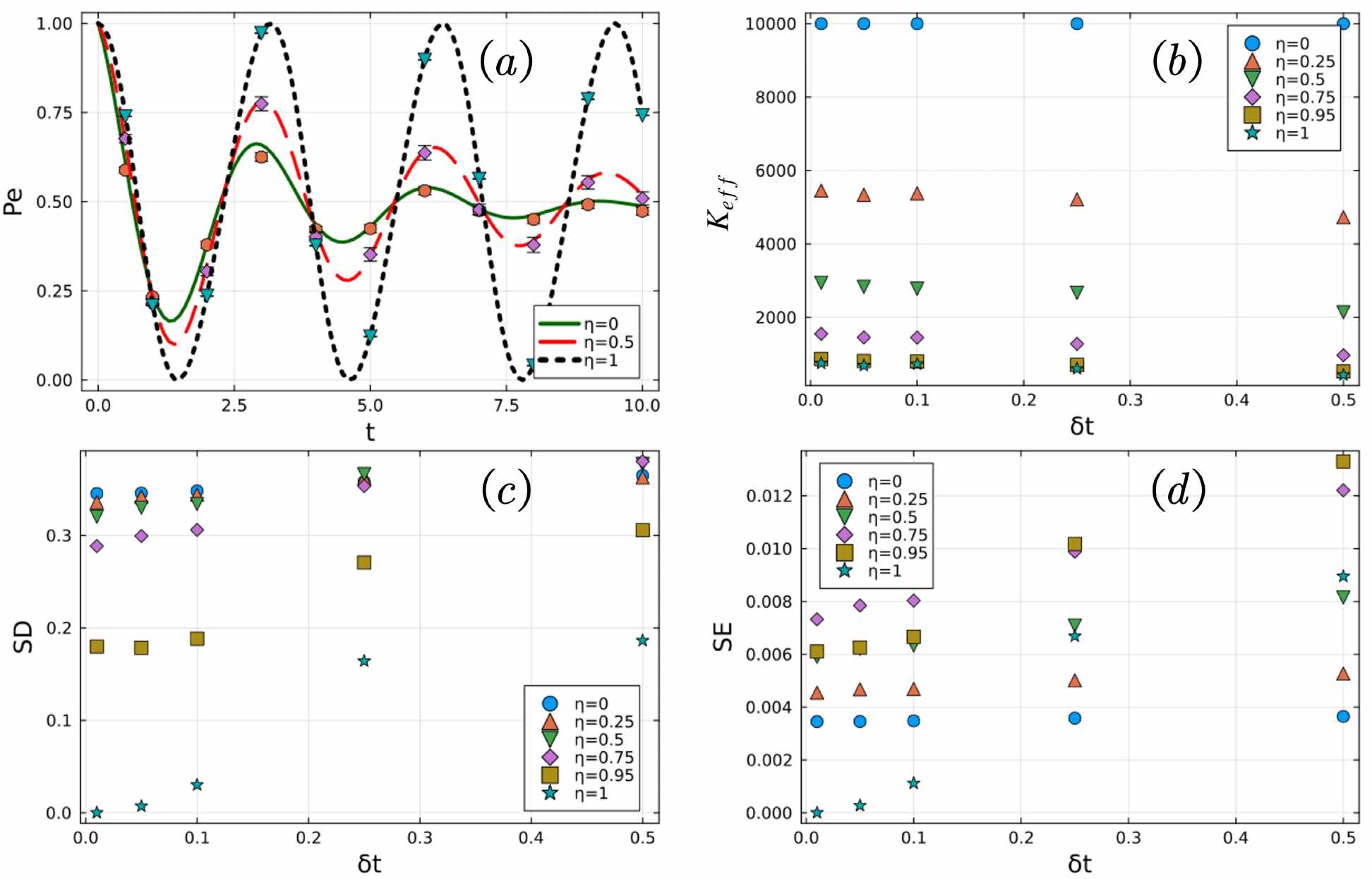}
\caption{ Simulation of two-level atom under monitoring spontaneous emission. We study the NLME evolution from $t=0$ to $10$. The atom is initialized in excited state,with parameters $(J,\gamma)=(1,0.5)$. (a) Time evolution of the excited state probability $P_e$ for different values of $\eta$. Digital simulation results (data points) from 1000 experimental rounds with time step of $\delta t =0.1$ are compared with exact solutions (lines) computed by the vectorization method. (b)-(d) The number of valid trajectories $K_{eff}$, standard deviation of $P_e$ (SD), and standard error of $P_e$ (SE) at time $t=10$ as a function of the time step $\delta t$ (=0.01,0.05,0.1,0.25,0.5) for different $\eta$ (=0.0,0.25,0.5,0.75,0.95,1.0). Each point comes from 10,000 experimental rounds.
}
\label{sfig::1}
\end{figure*}
%%%%%%%%%%%%%%%%%%%~~~~~~~~~~~%%%%%%%%%%%%%%%%

The time evolution of the excited-state probability $P_e=\Tr(|e\rangle \langle e| \rho )$ for different values of $\eta$ is presented in Fig.~\ref{sfig::1}(a). The solid curves represent the exact solutions obtained via the vectorization method, and the data points are the numerical results from our digital simulation, obtained with $\delta t=0.1$ and $K=1000$. The error bars represent the standard error (SE), which originates from the statistical sampling of valid trajectories. The SE is calculated as $SE=SD/\sqrt{K_{eff}}$, where the standard deviation (SD) is defined as:
%%%%%%%%%%%%%%%%%%%%%%%%
\begin{equation}
SD(t) = \sqrt{ \frac{1}{K_{\text{eff}}} \sum_{i=1}^{K_{\text{eff}}} \big[ P_e^i(t) - P_e^S(t) \big]^2 }.
\end{equation}
%%%%%%%%%%%%%%%%%%%%%%%%
Here, $P_e^i(t)$ is the excited-state probability in the i-th trajectory, and  $P_e^S(t)$ is the exact solution.

Furthermore, we numerically show the dependence of $K_{eff}$, $SD$, and $SE$ on $\delta t$ and $\eta$ at time $t=10$, as shown in Fig.~\ref{sfig::1}(b)-(d). For each data point, we perform $K=10,000$ rounds of numerical experiments. The numbers of effective trajectories are shown in Fig.~\ref{sfig::1}(b). The ratio $K_{eff}/K$ corresponds to the postselection success probability $P$, as given by Eq.~(9) in the main text, which implies that $K_{eff}$ follows an approximately exponential decay as a function of $\eta $. In the special case where $\eta=0$, we have $P=1$ and consequently $K_{eff}=K$.

The SD is influenced by two factors: (1) the non‑commutativity of the operators within $\delta t$, and (2) the stochastic probability distribution in Eq.~(4) in the main text. The former causes the SD to increase as $\delta t$ grows. While the latter, at a small $\delta t$, causes the SD to decrease as $\eta$ increases, due to the reduction in stochasticity. In particular, when $\eta=1$, the stochasticity vanishes completely. These features are clearly demonstrated in Fig.~\ref{sfig::1}(c).

Since $SE=SD/\sqrt{K_{eff}}$, the behavior of SE is governed by both SD and $K_{eff}$. As shown in Fig.~\ref{sfig::1}(d), SE increases with $\delta t$. This is because the increase in SD dominates the change in SE as $\delta t$ grows. In contrast, when $\eta$ increases, both $SD$ and $K_{eff}$ decrease. The competition between these two decreasing trends leads to the crossing of different types of points in Fig.~\ref{sfig::1}(d).

%%%%%%%%%%%%%%%%%%%%%%%%%%%%%%%%%%%%%%%%%%%%%%%%%%%%%%%%%%%%%%%%%%%%%%%%%%%%%%%%%%%%%%%%%%%%%%%%%%%%%%%%%%%%%%%%%%%%%%%%
%%%%%%%%%%%%%%%%%%%%%%%%%%%%%%%%%%%%%%%%%%%%%%%%%%%%%%%%%%%%%%%%%%%%%%%%%%%%%%%%%%%%%%%%%%%%%%%%%%%%%%%%%%%%%%%%%%%%%%%%
\subsection{S3.\ Simulation of many-body localization in open system and the postselected skin effect}

As stated in the Introduction, many novel theoretical frameworks based on specially constructed Lindbladians have been proposed in recent years. While these approaches reveal rich physical phenomena, they are frequently challenging to implement experimentally. Moreover, for problems of high computational complexity, purely theoretical analyses often fail to provide a comprehensive understanding. These challenges clearly demonstrate the significance of quantum-circuit-based simulation as an efficient and controllable platform for studying open quantum dynamics. 

Here, we present two representative examples: many-body localization in open system~\cite{Yusipov2017,Vakulchyk2018,YLiu2024} and the postselected skin effect. The postselected skin effect represents a class of phenomena that are theoretically understood but difficult to realize on analog platforms. \textcolor{red}{  In contrast, many-body localization in open quantum systems remains a problem without a definitive theoretical conclusion.}

\textcolor{red}{ Our disorder model extends the frameworks presented in Ref.~\cite{YLiu2024}, where dissipative terms can induce effective particle–particle interactions. However, their analyses were restricted to the single-particle sector, focusing solely on Anderson localization in open systems. Here, we present a digital simulation of a half-filled many-body model, in which features of localization can be observed.}

\textcolor{red}{We emphasize that a rigorous theoretical treatment of many-body localization in open systems, which inherently requires large-scale simulations, remains extremely challenging due to the limitations of classical computational power. Our results of simulation therefore serve as a small-scale, proof-of-principle demonstration, highlighting the value of digital quantum simulation in the Noisy Intermediate-Scale Quantum (NISQ) era for exploring theoretically difficult open-system problems.}

%%%%%%%%%%%%%%%%%%%%%%%%%%%%%%%%%%%%%%%%%%%%%%%%%%%%%%%%%%%%%%%%%%%%%%%%%%%%%%%%%%%%%%%%%%%%%%%%%%%%%%%%%%%%%%%%%%%%%%%%
\subsubsection{Digital simulation of many-body localization in open system}

Consider an \textit{L}-site half-filled spin chain with the LME,
%%%%%%%%%%%%%%%%%%%%%%%%
\begin{equation}
\begin{split}~\label{Local}
& \frac{d}{dt} \rho = -i[H,\rho] + \gamma \sum_{l=1}^{L-1} \left( -\frac{1}{2} \{ L_{l}^{\dag} L_{l}, \rho \} +  L_{l} \rho L_{l}^{\dag} \rho \right),\\
& H=J\sum_{l=1}^{L-1} \left( \sigma_l^+ \sigma_{l+1}^- + {\rm H.c.} \right) + \sum_{l=1}^{L} V \cos (2\pi \omega l) \sigma_l^z, 
\end{split}
\end{equation}
%%%%%%%%%%%%%%%%%%%%%%%%
where $\sigma_l^\mu (\mu=z,+,-) $ is the spin-1/2 operator at the site $l$, and $\omega = (\sqrt{5} -1)/2$ indicates the quasicrystal potential. The local dissipation operator takes the following form:
%%%%%%%%%%%%%%%%%%%%%%%%
\begin{equation}~\label{Ll}
L_l=\frac{1}{2}(\sigma_l^+ + e^{i\alpha}\sigma_{l+1}^+)(\sigma_l^- + e^{i\beta}\sigma_{l+1}^-),
\end{equation}
%%%%%%%%%%%%%%%%%%%%%%%%
which represents the phase-changing effect between neighboring spins. This kind of dissipation operator exhibits novel phenomena in theoretical studies, such as entangled state preparation~\cite{BKraus2008,SDiel2008,PSchindler2013}, Anderson localization~\cite{Yusipov2017,YLiu2024}, and the skin effect in open quantum systems~\cite{Hugo2025,Wang2022,Feng2023,ZCLiu2024}.

\textcolor{red}{Note that the evolution from $\exp(-\frac{1}{2} \gamma L_l^\dag L_l \delta t)$ indicates this model is an intrinsic many-body problem due to the $\hat{n}_l \, \hat{n}_{l+1}$ interaction ($\hat{n}_l :=\sigma_{l}^+ \sigma_{l}^- $) existing in}
%%%%%%%%%%%%%%%%%%%%%%%%
\begin{equation}~\label{SLl}
L_l^\dag L_l = \frac{1}{2}\left(\hat{n}_l + \hat{n}_{l+1} + e^{i\beta} \sigma^{+}_l \sigma^{-}_{l+1} + e^{-i\beta} \sigma^{+}_{l+1} \sigma^{-}_l + \textcolor{red}{ [\cos(\alpha + \beta) -1 ]  \hat{n}_l  \hat{n}_{l+1}  } \right),
\end{equation}
%%%%%%%%%%%%%%%%%%%%%%%%
except for $\cos(\alpha + \beta) = 1$.

We simulate the dynamics by the 1-dilation method as shown in Fig.~(2) b in the main text. The core gate $U_l$, corresponding to Eq.~(11) in the main text, is given by
%%%%%%%%%%%%%%%%%%%%%%%%
\begin{equation}
U_l = \left(
\begin{matrix}
B_l & \tilde{A}_l \\
A_l & -B_{l}^\dag
\end{matrix} \right),
\end{equation}
%%%%%%%%%%%%%%%%%%%%%%%%
where
%%%%%%%%%%%%%%%%%%%%%%%%
\begin{equation}\label{SAB}
\begin{split}
& A_l = \sqrt{1-\gamma \delta t L_l^\dag L_l}
=\left(\begin{matrix} \sqrt{1-\gamma \delta t \cos^2 (\frac{\alpha+\beta}{2})} & 0 & 0 & 0 \\  0 & (1+\sqrt{1-\gamma \delta t})/2 & (\sqrt{1-\gamma \delta t}-1)e^{i\beta}/2 & 0 \\ 0 & (\sqrt{1-\gamma \delta t}-1)e^{-i\beta}/2 & (1+\sqrt{1-\gamma \delta t})/2 & 0 \\ 0 & 0 & 0 & 1 \end{matrix}\right) \\
& \tilde{A}_l = \sqrt{1-\gamma \delta t L_l L_l^\dag}
=\left(\begin{matrix} \sqrt{1-\gamma \delta t \cos^2 (\frac{\alpha+\beta}{2})} & 0 & 0 & 0 \\  0 & (1+\sqrt{1-\gamma \delta t})/2 & (\sqrt{1-\gamma \delta t}-1)e^{-i\alpha}/2 & 0 \\ 0 & (\sqrt{1-\gamma \delta t}-1)e^{i\alpha}/2 & (1+\sqrt{1-\gamma \delta t})/2 & 0 \\ 0 & 0 & 0 & 1 \end{matrix}\right) \\
& B_l = \sqrt{\gamma \delta t} \, L_l
=\frac{\sqrt{\gamma \delta t}}{2} \left(\begin{matrix} 1+e^{i(\alpha + \beta)} & 0 & 0 & 0 \\  0 & 1 & e^{i\beta} & 0 \\ 0 & e^{i\alpha} & e^{i(\alpha + \beta)} & 0 \\ 0 & 0 & 0 & 0 \end{matrix}\right).
\end{split}
\end{equation}
%%%%%%%%%%%%%%%%%%%%%%%%

\begin{figure*}[htbp]\centering
\includegraphics[width=15cm]{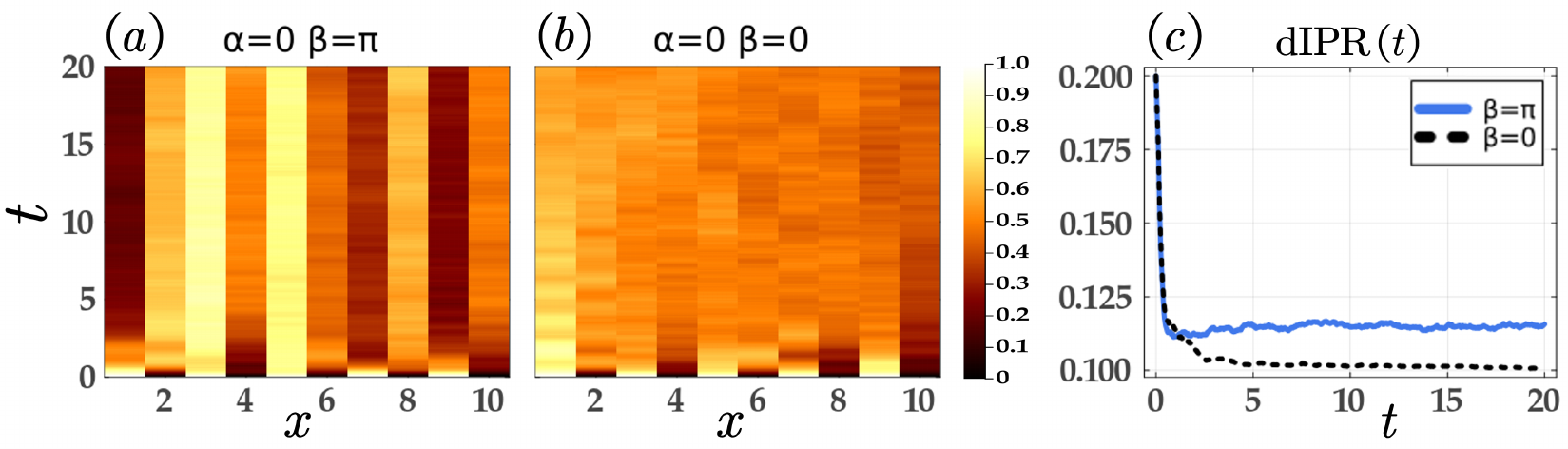}
\caption{
Simulation of localization and thermalization in an open system. The probability distribution of the spin-up state in a 10-site chain evolves from the initial state $|\uparrow  \downarrow \uparrow \cdots \downarrow \rangle$. The dynamics is govern by the LME in Eq.~(\ref{Local}) with $J=1$, $V=2$, $\gamma=1$ and $\eta=0$. The jump operators take the form as Eq.~(\ref{Ll}) with $\alpha=0$, $\beta=\pi$ in (a) and  $\alpha=0$, $\beta=0$ in (b). We simulate the evolution by the 1-dilation method with $\delta t=0.01$ and an average over 100 trajectories. The results show that the system is localized in (a) and thermalized in (b). The evolution curves of dIPR in both (a) and (b) are shown in (c).
}
\label{sfig::2}
\end{figure*}

In Ref.~\cite{YLiu2024}, Anderson localization is demonstrated in the single-particle space for parameters $\alpha=0$ and $\beta=\pi$. Here, we extend this case to many-particle systems, showing that localization features persist even in such cases. As illustrated in Fig.~\ref{sfig::2}(a), starting from the initial state $|\uparrow  \downarrow \uparrow  \downarrow \cdots \rangle$ , the probability distribution of the spin-up state, $\langle \hat{n}_l \rangle$, exhibits clear localization signatures. As a comparison, in the case of $\alpha=\beta=0$, the system fully thermalizes, as shown in Fig.~\ref{sfig::2}(b). To quantify the difference between the two cases, we focus on the dynamic inverse participation ratio (dIPR)~\cite{TLi2022}, 
%%%%%%%%%%%%%%%%%%%%%%%%
\begin{equation}
{\rm dIPR}(t)=\left(  \sum_{l=1}^L \langle \hat{n}_l (t) \rangle^2  \right) / (\sum_{l=1}^L \langle \hat{n}_l (t) \rangle)^2,
\end{equation}
%%%%%%%%%%%%%%%%%%%%%%%%
which is near $1/L$ for extended states and has a larger value for localized states. The curves of dIPR are shown in Fig.~\ref{sfig::2}(c), where the value for $\beta=\pi$ is clearly larger than that for $\beta=0$.

%%%%%%%%%%%%%%%%%%%~~~~~~~~~~~%%%%%%%%%%%%%%%%%%%
%%%%%%%%%%%%%%%%%%%~~~~~~~~~~~%%%%%%%%%%%%%%%%

%%%%%%%%%%%%%%%%%%%%%%%%%%%%%%%%%%%%%%%%%%%%%%%%%%%%%%%%%%%%%%%%%%%%%%%%%%%%%%%%%%%%%%%%%%%%%%%%%%%%%%%%%%%%%%%%%%%%%%%%
\subsubsection{Digital simulation of postselected skin effect}

Consider an \textit{L}-site spin chain with the NLME,
%%%%%%%%%%%%%%%%%%%%%%%%
\begin{equation}
\frac{d}{dt} \rho = -i[H,\rho] + \gamma \sum_{l=1}^{L-1} \left(-\frac{1}{2} \{ L_{l}^{\dag} L_{l}, \rho \} + (1-\eta) L_{l} \rho L_{l}^{\dag} + \eta  \langle L_{l}^{\dag} L_{l} \rangle \rho \right),
\end{equation}
%%%%%%%%%%%%%%%%%%%%%%%%
where $H=J\sum_{l=1}^{L-1} \left( \sigma_l^+ \sigma_{l+1}^- + {\rm H.c.} \right)$  and the $L_l$ is the same as it in Eq.~(\ref{SLl}).
%%%%%%%%%%%%%%%%%%%%%%%%

We simulate the dynamics by the 2-dilation method as shown in Fig.~(2) a in the main text. The core gate $U_l$, corresponding to Eq.~(7) in the main text, is given by
%%%%%%%%%%%%%%%%%%%%%%%%
\begin{equation}\label{SU}
U_l = \left(
\begin{matrix}
C_l & B_l &  \tilde{A}_l & 0 \\ B_l & -C_l & 0 & \tilde{A}_l \\
A_l & 0 & -C_l^\dag & -B_l^\dag \\ 0 & A_l & -B_l^\dag & C_l^\dag
\end{matrix} \right) ,
\end{equation}
%%%%%%%%%%%%%%%%%%%%%%%%
where $A_l$ and $\tilde{A}_l$ are the same as those in Eq.~(\ref{SAB}), and
%%%%%%%%%%%%%%%%%%%%%%%%
\begin{equation}\label{SABC}
\begin{split}
& B_l = \sqrt{(1-\eta)\gamma \delta t} \, L_l
=\frac{\sqrt{(1-\eta)\gamma \delta t}}{2} \left(\begin{matrix} 1+e^{i(\alpha + \beta)} & 0 & 0 & 0 \\  0 & 1 & e^{i\beta} & 0 \\ 0 & e^{i\alpha} & e^{i(\alpha + \beta)} & 0 \\ 0 & 0 & 0 & 0 \end{matrix}\right)\\
& C_l = \sqrt{\eta \gamma \delta t} \, L_l
=\frac{\sqrt{\eta \gamma \delta t}}{2}\left(\begin{matrix} 1+e^{i(\alpha + \beta)} & 0 & 0 & 0 \\  0 & 1 & e^{i\beta} & 0 \\ 0 & e^{i\alpha} & e^{i(\alpha + \beta)} & 0 \\ 0 & 0 & 0 & 0 \end{matrix}\right).
\end{split}
\end{equation}
%%%%%%%%%%%%%%%%%%%%%%%%

%%%%%%%%%%%%%%%%%%%~~~~~~~~~~~%%%%%%%%%%%%%%%%%%%
\begin{figure*}[htbp]\centering
\includegraphics[width=15cm]{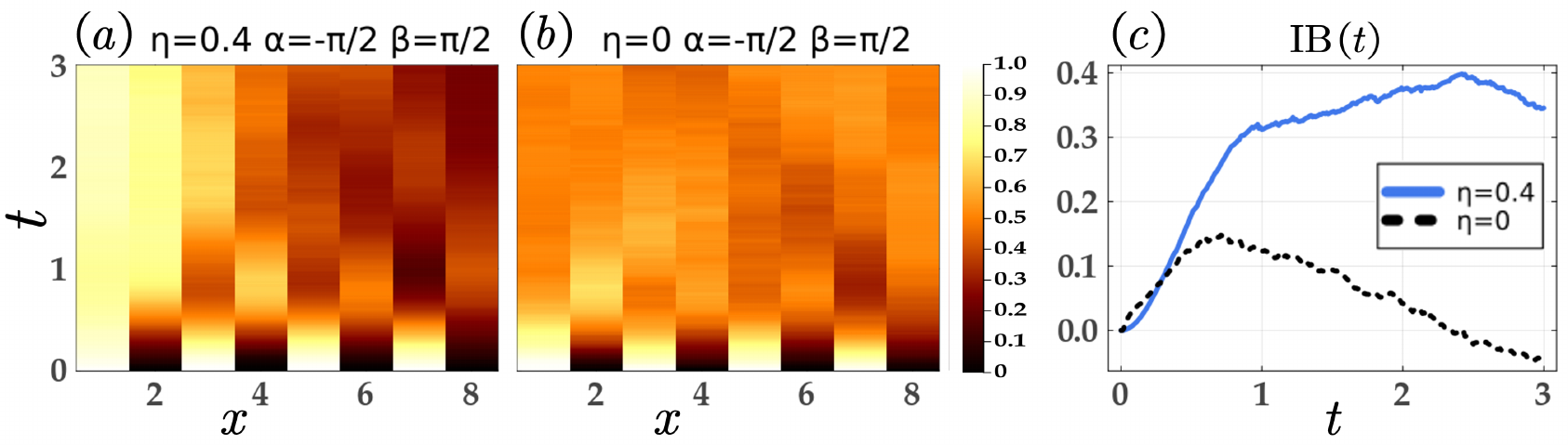}
\caption{
Simulation of the postselected skin effect. The probability distribution of the spin-up state in an 8-site chain evolves from the initial state $|\uparrow  \downarrow \uparrow \cdots \downarrow \rangle$. The dynamics is governed by the NLME in Eq.~(\ref{Local}) with $J=1$, $V=0$, and $\gamma=2$. The jump operators take the form of Eq.~(\ref{Ll}) with $\alpha=-\pi /2$, $\beta=\pi /2$. When postselection is applied, the spin-up state tends to move toward the left as shown in (a) with $\eta=0.4$. As a comparison, for $\eta=0$, the system is thermalized in (b). The results are simulated using the 2-dilation method with $\delta t=0.01$ and an average over 90 valid trajectories for $\eta=0.4$ and 100 trajectories for $\eta=0$. The evolution curves of IB in both (a) and (b) are shown in (c).
}
\label{sfig::3}
\end{figure*}
%%%%%%%%%%%%%%%%%%%~~~~~~~~~~~%%%%%%%%%%%%%%%%

When $\alpha=\pi/2$, $\beta=\pi/2$, and $\eta \ne 0$, the steady state is expected to exhibit the postselected skin effect~\cite{Hugo2025}. We simulate this dynamics of an 8-site chain from the initial state $|\uparrow  \downarrow \uparrow \cdots \downarrow \rangle$. When $\eta=0.4$, the postselection causes the probability distribution of the spin-up state to become concentrated toward the left, as shown in Fig.~\ref{sfig::3}(a). As a comparison, in the case of $\eta=0$, the system tends to thermalize, as shown in Fig.~\ref{sfig::3}(b). To quantify the difference between the two cases, we define the imbalance between left and right part of the chain by
%%%%%%%%%%%%%%%%%%%%%%%%
\begin{equation}
{\rm IB}(t)=\left(  \sum_{l=1}^{L/2} \langle \hat{n}_l (t) \rangle -  \sum_{l=L/2+1}^{L} \langle \hat{n}_l (t) \rangle \right) / (\sum_{l=1}^L \langle \hat{n}_l (t) \rangle).
\end{equation}
%%%%%%%%%%%%%%%%%%%%%%%%
As shown in Fig.~\ref{sfig::3}(c), the ${\rm IB}$ tends to $0$ for $\eta=0$ and ${\rm IB} \ne 0$ for $\eta=0.4$, which implys the postselected skin effect.

% Additionally, we would like to mention that when $\eta\ne0$, the simulation of the NLME is a probabilistic realization, which is still challenged by the success probability approaching zero. This is reflected in the fact that we obtain 90 valid trajectories from 300000 trials in the simulation for $\eta=0.4$. However, the simulation of the standard LME is a deterministic realization. Therefore, we obtain 100 valid trajectories just 100 trials of simulation for $\eta=0$

\end{document}